\documentclass[prl,twocolumn,superscriptaddress,floatfix,showpacs,amsmath,amssymb]{revtex4}

\usepackage{graphicx}
\usepackage{dcolumn}
\usepackage{bm}

\begin{document}

\title{Measurements of $\vec{\text{H}}\vec{\text{D}}(\vec{\gamma},\pi)$
and Implications for Convergence of the GDH Integral}

\author{S.~Hoblit}\email[Corresponding author: ]{hoblit@bnl.gov}
\affiliation{Deptartment of Physics, University of Virginia, Charlottesville, VA 22901}
\affiliation{Physics Department, Brookhaven National Laboratory, Upton, NY 11973}

\author{A.~M.~Sandorfi}\email[Corresponding author: ]{sandorfi@jlab.org}
\affiliation{Physics Department, Brookhaven National Laboratory, Upton, NY 11973}

\author{K.~Ardashev}
\affiliation{Deptartment of Physics, University of Virginia, Charlottesville, VA 22901}
\affiliation{Deptartment of Physics, University of South Carolina, Columbia, SC 29208}

\author{C.~Bade}
\affiliation{Department of Physics, Ohio University, Athens OH 45701}

\author{O.~Bartalini}
\affiliation{Universit\`{a} di Roma ``Tor Vergata'' and INFN-Sezione di Roma2, Rome, Italy}

\author{M.~Blecher}
\affiliation{Physics Deptartment, Virginia Polytechnic Inst.\ \& State Univ., Blacksburg, VA 24061}

\author{A.~Caracappa}
\affiliation{Physics Department, Brookhaven National Laboratory, Upton, NY 11973}

\author{A.~D'Angelo}
\affiliation{Universit\`{a} di Roma ``Tor Vergata'' and INFN-Sezione di Roma2, Rome, Italy}

\author{A.~d'Angelo}
\affiliation{Universit\`{a} di Roma ``Tor Vergata'' and INFN-Sezione di Roma2, Rome, Italy}

\author{R.~Di~Salvo}
\affiliation{Universit\`{a} di Roma ``Tor Vergata'' and INFN-Sezione di Roma2, Rome, Italy}

\author{A.~Fantini}
\affiliation{Universit\`{a} di Roma ``Tor Vergata'' and INFN-Sezione di Roma2, Rome, Italy}

\author{C.~Gibson}
\affiliation{Deptartment of Physics, University of South Carolina, Columbia, SC 29208}

\author{H.~Gl\"{u}ckler}
\affiliation{Forschungszentrum J\"{u}lich GmbH, D-52425 J\"{u}lich, Germany}

\author{K.~Hicks}
\affiliation{Department of Physics, Ohio University, Athens OH 45701}

\author{A.~Honig}
\affiliation{Deptartment of Physics, Syracuse University, Syracuse, NY 13210}

\author{T.~Kageya}
\affiliation{Physics Deptartment, Virginia Polytechnic Inst.\ \& State Univ., Blacksburg, VA 24061}

\author{M.~Khandaker}
\affiliation{Norfolk State University, Norfolk and Jefferson Lab, Newport News, VA 23606}

\author{O.~C.~Kistner}
\affiliation{Physics Department, Brookhaven National Laboratory, Upton, NY 11973}

\author{S.~Kizilgul}
\affiliation{Department of Physics, Ohio University, Athens OH 45701}

\author{S.~Kucuker}
\affiliation{Deptartment of Physics, University of Virginia, Charlottesville, VA 22901}

\author{A.~Lehmann}
\affiliation{Deptartment of Physics, University of South Carolina, Columbia, SC 29208}

\author{M.~Lowry}
\affiliation{Physics Department, Brookhaven National Laboratory, Upton, NY 11973}

\author{M.~Lucas}
\affiliation{Department of Physics, Ohio University, Athens OH 45701}

\author{J.~Mahon}
\affiliation{Department of Physics, Ohio University, Athens OH 45701}

\author{L.~Miceli}
\affiliation{Physics Department, Brookhaven National Laboratory, Upton, NY 11973}

\author{D.~Moricciani}
\affiliation{Universit\`{a} di Roma ``Tor Vergata'' and INFN-Sezione di Roma2, Rome, Italy}

\author{B.~Norum}
\affiliation{Deptartment of Physics, University of Virginia, Charlottesville, VA 22901}

\author{M.~Pap}
\affiliation{Forschungszentrum J\"{u}lich GmbH, D-52425 J\"{u}lich, Germany}

\author{B.~Preedom}
\affiliation{Deptartment of Physics, University of South Carolina, Columbia, SC 29208}

\author{H.~Seyfarth}
\affiliation{Forschungszentrum J\"{u}lich GmbH, D-52425 J\"{u}lich, Germany}

\author{C.~Schaerf}
\affiliation{Universit\`{a} di Roma ``Tor Vergata'' and INFN-Sezione di Roma2, Rome, Italy}

\author{H.~Str\"{o}her}
\affiliation{Forschungszentrum J\"{u}lich GmbH, D-52425 J\"{u}lich, Germany}

\author{C.~E.~Thorn}
\affiliation{Physics Department, Brookhaven National Laboratory, Upton, NY 11973}

\author{C.~S.~Whisnant}
\affiliation{James Madison University, Harrisonburg, VA 22807}

\author{K.~Wang}
\affiliation{Deptartment of Physics, University of Virginia, Charlottesville, VA 22901}

\author{X.~Wei}
\affiliation{Physics Department, Brookhaven National Laboratory, Upton, NY 11973}

\collaboration{LSC Collaboration}
\noaffiliation

\date{\today}

\begin{abstract}

We report new measurements of inclusive $\pi$ production from frozen-spin HD
for polarized photon beams covering the $\Delta$(1232) resonance.
These provide data simultaneously on both H and D with nearly complete
angular distributions of the spin-difference cross sections entering
the Gerasimov-Drell-Hearn (GDH) sum rule. Recent results from Mainz and
Bonn exceed the GDH prediction for the proton by 22 $\mu$b, suggesting as
yet unmeasured high-energy components. Our $\pi^{\circ}$ data 
reveal a different angular dependence than assumed in Mainz analyses
and integrate to a value that is 18 $\mu$b lower, suggesting
a more rapid convergence. Our results for deuterium
are somewhat lower than published data, considerably more
precise and generally lower than available calculations.

\end{abstract}


\maketitle

In 1966 three sets of authors, Gerasimov \cite{Gera}, Drell and Hearn \cite{drell} and Hosoda
and Yamamoto \cite{hosoda} independently derived a sum rule for the anomalous magnetic
moment ($\kappa$) of spin $S=1/2$ particles in terms of the energy-weighted difference
between total photon reaction cross sections in entrance channel
states with parallel (\textit{P}) and anti-parallel (\textit{A}) photon and target
spin alignments, 
\begin{equation}
\int_{\omega_{0}}^{\infty}\frac{\sigma_{P} - \sigma_{A}}{\omega}d\omega =
4S\pi^{2}\alpha\left(\frac{\kappa}{m}\right)^2 .
\label{GDHeq}
\end{equation}
In recent literature this relation for $S=1/2$ nucleons has been referred to
as the \textit{GDH sum rule}. Hosoda and Yamamoto also showed that the same relation
holds for spin $S=1$ nuclei, such as the deuteron \cite{hos2}. This expression follows
from a Gell-mann\textendash Goldberger\textendash Thirring dispersion relation for the forward
elastic (Compton) amplitude \cite{gellman}, provided that the spin-flip Compton
amplitude vanishes at high energy at least as fast as $1/ln(\omega)$. Because of
the latter requirement, this sum rule is not fundamental, in that no
underlying theory falls if it is violated. Rather, convergence of the
above integral to a value different from the right side of eqn.~\ref{GDHeq} would
reveal an interesting property of a very high-energy process.

For the proton and deuteron, the right hand side of eqn.~\ref{GDHeq} reduces to
204 $\mu b$ and 0.7 $\mu b$, respectively. Recently, a collaboration from Mainz and
Bonn has experimentally checked the GDH
sum rule for the proton \cite{dutz}, and at least consistency with calculations for
the deuteron over a limited energy range \cite{ahrens06}. Their proton measurements
spanned the energy range from 0.2 to 2.9 GeV and yielded
$254 \pm 5 \pm 12$ $\mu$b, exceeding the sum rule expectation. Multipole analyses
such as SAID \cite{arndt} or MAID \cite{drechsel} agree that a contribution of
$-28 \mu b$ is expected from the near threshold region below 0.2 GeV. This
would require an as yet unmeasured $-22 \mu b$ from high
energies to restore agreement with the expectations of eqn.~\ref{GDHeq}, which is
possible since some negative contributions have been suggested by
Regge models \cite{simula,helbing}.

We report here new measurements of inclusive $\pi$ photo-production from a
polarized HD target, spanning a range of polarized photon energies covering
the P$_{33}$ $\Delta$ resonance. The experiments were performed 
at the Laser Electron Gamma Source (LEGS) at Brookhaven National
Laboratory with tagged circular polarized $\gamma$-rays between 190 and 420 MeV.
The general
characteristics of the LEGS photon beams are discussed in ref \cite{legs}.
Here the photon polarization averaged between 60\% to 99\% 
and was cycled between left and right circular states
at randomly chosen times averaging every few minutes. 

The polarized target consisted of solid hydrogen-deuteride (HD), held
in a frozen-spin state. The material was condensed in a variable temperature
cryostat, where the NMR polarization monitoring system was calibrated
at 2 K, transferred to a dilution refrigerator for polarization at $\sim$15 mK
and 15 Tesla, held there for typically 3 months to reach the frozen-spin
state, and finally transferred to an In-Beam Cryostat (IBC) operating at
0.3 K, where a thin 0.9 Tesla solenoid maintained the H and D orientations.
The polarization cycle will be detailed in a separate publication.
Some aspects are discussed in \cite{andy}.
Data were collected during two running periods in Fall 2004 and Spring 2005, the
first emphasizing H polarization, with initial polarizations of $P(H) = 0.59$
and $P(D) = 0.07$, and the second using increased D polarization
following an RF transfer of spin between H and D, with $P(H) = 0.32$ and $P(D) = 0.33$.
Mid-way through each
period the H polarization was flipped with an RF transition. 
This produced four distinct data blocks with differing target polarizations,
during which the in-beam spin relaxation times for polarized H and D
ranged from 7 to 15 months.
The polarization was monitored frequently with a cross-coil NMR system within
the IBC \cite{thorn}. 

For these measurements, pions were detected in a large \textit{Spin Asymmetry}
(SASY) calorimeter. An array of 432 NaI(Tl) detectors, an \textit{XBOX},
surrounded the target covering laboratory (Lab) angles
from $45^\circ$ to $135^\circ$. A cylindrical array of plastic 
neutron detectors was positioned between the XBOX 
and the IBC. (Results for exclusive channels 
will be discussed elsewhere.) A forward wall consisting
of 31 cm of plastic scintillators, backed by an array of 176
Pb-Glass crystals detected reaction products 
at Lab angles between $10^\circ$ and $40^\circ$. The configuration of these detectors
was optimized for neutral pions, with either two decay photons detected
in the XBOX or one in the XBOX and the other in the forward wall.
This provided nearly complete coverage for $\pi^\circ$ detection.

\begin{figure}
\includegraphics[width=8.0cm]{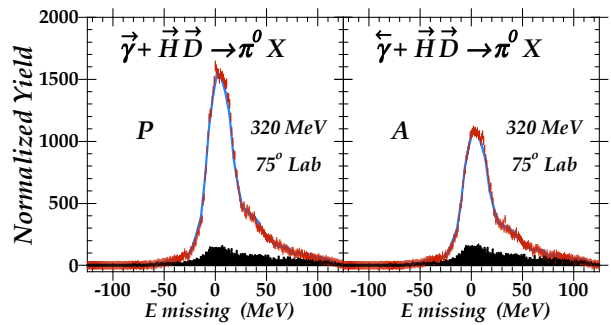}
\caption{\label{fig:pidif} Differences between 2-body kinematics and the measured $\pi^\circ$ energy 
are shown in red, for parallel (left) and anti-parallel (right) beam and target
spin alignments. Simulated energy differences are shown as the solid
(blue) curves. Empty cell contributions are shaded in black.}
\end{figure}

Two-pion production is negligible throughout our energy
range \cite{krusche}. As a result, spectra at a fixed angle and tagged energy are
dominated by 2-body (from H) or quasi-2-body (from D) kinematics. 
The energy of reconstructed
neutral pions is compared
to the 2-body expectation in Fig.~\ref{fig:pidif} for one of 10
angle bins, 17 tagged energy bins and 4 target polarization groups.
The simulated response (blue curve) is in excellent agreement.
The spin asymmetry is evident in the left and right
panels, which show yields for parallel and anti-parallel beam and target
spin alignments. Charged pion spectra are very similar.

The only unpolarizable nucleons in the target are found in a mesh of 50 $\mu m$
Al wires used to conduct away heat during polarization and
in pCTFE (C$_2$ClF$_3$) windows of the target cell. Their contributions are
determined through empty cell measurements (black area in Fig.~\ref{fig:pidif}).

We discuss here inclusive $\pi$ production, integrated over azimuthal angles, for
which the differential cross section from polarized HD can be written as \cite{andy,fix},
\begin{equation}
d\sigma(\theta,E_{\gamma}) = d\sigma_{0}^{HD} - P_{\gamma}^{c}P_{H}\hat{E}_{H} - 
P_{\gamma}^{c}P_{D}^{V}\hat{E}_{D} + \sqrt{1/2}P_{D}^{T}\hat{T}_{20}^{0} ,
\label{gensig}
\end{equation}
where $P_{\gamma}^{c}$ is the circular beam polarization, $P_H$ is the hydrogen polarization,
$P_D^V$ and $P_D^T$ are the deuteron vector and tensor polarizations, respectively, and
a subscript zero (0) denotes an unpolarized cross section. Here we have
designated the numerator of a spin asymmetry with a carat, so that
$\hat{E}_H = d\sigma_0^H E_H = 1/2\left[ d\sigma^{H}(A) - d\sigma^{H}(P)\right]$,
$\hat{E}_D$ is the corresponding quantity for deuterium and  
$\sqrt{1/2}\hat{T}_{20}^0 = \sqrt{1/2}d\sigma^{D}_{0}T_{20}^0 = 1/2\left[ d\sigma^{D}(A)
+ d\sigma^{D}(P) - 2d\sigma_0^D \right]$
is the deuteron tensor observable, following the convention of \cite{fix}.

The data set consists of four distinct blocks
with different target polarizations, each containing roughly equal amounts
of data with right and left circular photon polarization. These eight
data groups overdetermine the four observables of eqn.~\ref{gensig}. 
Fits varying $\hat{T}_{20}^0$ produced at most few percent changes in
$d\sigma_{0}^{HD}$, compared to fixing $\hat{T}_{20}^0$ to zero, and no
perceptible changes to $\hat{E}_H$ and $\hat{E}_D$. 
Here we focus on results of fits with $\hat{T}_{20}^0$ fixed to zero.

\begin{figure}
\includegraphics[width=8.0cm,height=5.5cm]{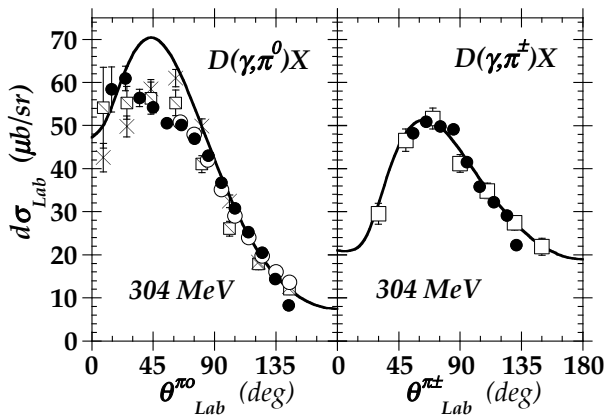}
\caption{\label{fig:unpold} Unpolarized cross sections (solid circles) for
D$(\gamma,\pi^{\circ})$X, left, and D$(\gamma,\pi^{\pm})$X, right, at $E_\gamma = 304$ MeV,
deduced by subtracting SAID(FA07k) predictions \cite{arndt} for p$(\gamma,\pi)$ from
fitted HD results. For the $\pi^\circ$ channel, LEGS data from a liquid
D$_2$ target are shown as open circles, while crosses and hatched-boxes
are from \cite{siodlaczek} and \cite{krusche}. For the $\pi^\pm$ channel, open boxes
are constructed from $\pi^{-}$pp \cite{benz} and the $\pi^{-}/\pi^{+}$ ratio data of \cite{fujii}.
The curves are calculations from \cite{fix}.}
\end{figure}

Sample angular distributions of the unpolarized cross section at the
peak of the $\Delta$(1232) are shown in Fig.~\ref{fig:unpold} (solid circles). To compare
with other available deuteron data we have subtracted the well-known
proton cross sections as parameterized by SAID(FA07k) \cite{arndt}. Here we
show results for the Fall'04 data for which the deuteron tensor
polarization was negligible.

The normalization scale was checked by comparing D$(\gamma,\pi^{\circ})$X
cross sections to data collected with the same detector array using
a liquid D$_2$ target of known length (open circles in Fig.~\ref{fig:unpold}). The
Fall'04 target was grown slowly and its length agreed with
that expected from the known amount of
HD gas.  The Spring'05 target was grown rapidly and its
cross section scale was normalized to the
Fall'04 D$(\gamma,\pi^{\circ})$X by fitting to the interval
$110^{\circ} \leq \theta_{Lab}^{\pi} \leq 150^{\circ}$ where
both our fits and the calculations of \cite{fix} agree that $\hat{T}_{20}^0$ is negligible.

\begin{figure}
\includegraphics[width=8.0cm,height=10cm]{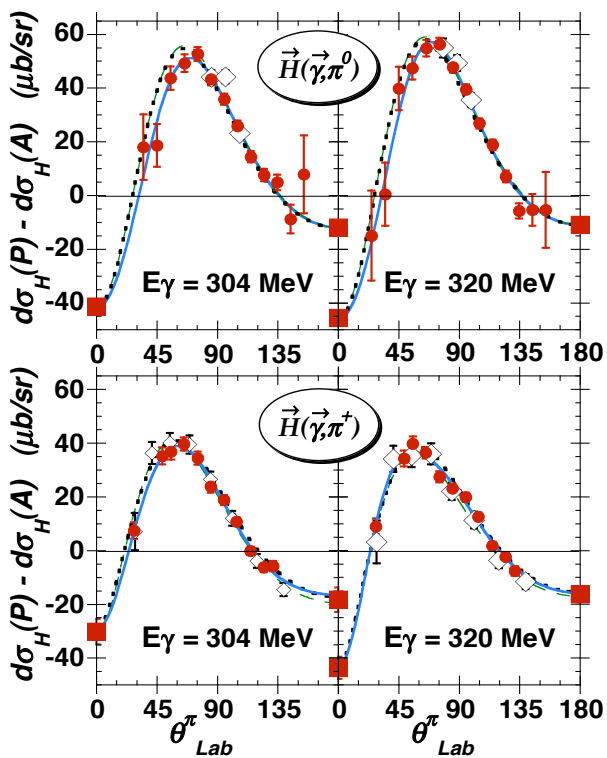}
\caption{\label{fig:angd} Angular dependence of the $[P-A]$ spin-difference
cross section for polarized H at beam energies near the $\Delta$ peak.  The
full data are solid (red) circles.
Unpolarized limits (solid red squares) at $0^{\circ}$ and $180^{\circ}$ are the means
of SAID \cite{arndt} and MAID \cite{drechsel}. Open diamonds are
from Mainz \cite{ahrens04}, interpolated to these energies. Predictions
from SAID and MAID are dotted (black) and dashed (green)
curves, respectively. Solid (blue) lines show Legendre fits to our data.}
\end{figure}

Differential spin-difference cross sections,
$[d\sigma^{H}(P) - d\sigma^{H}(A)] = -2\hat{E}_H$, for polarized H
are shown in Fig.~\ref{fig:angd} as solid circles for energies near the peak
of the $\Delta$. Since the pion has zero spin, at $0^{\circ}$ and
$180^{\circ}$ the H spin
difference reduces to $-2d\sigma^{H}_0$ . For these angles, the mean of
SAID(FA07k) \cite{arndt} and MAID(2007) \cite{drechsel} were used
(solid squares). The Mainz H-Butanol results
for $\pi^+$ are in very good
agreement with the present data, both
here and at other energies. SAID and MAID multipole predictions,
which include the Mainz data in their fits, reproduce the angular
dependence of the $\pi^+$  spin difference. The Mainz $\pi^{\circ}$ differential spin difference
data are again in good agreement with our results, although they
have a very limited angular range \cite{ahrens04}. However, forward of $80^{\circ}$ in
the laboratory, our spin-difference results drop below the multipole
prediction of SAID and MAID. This trend occurs mainly near the $\Delta$ peak.
At energies 40 MeV higher or lower, SAID and MAID $\pi^{\circ}$ 
predictions are quite close to our data.

\begin{figure}[ht!]
\includegraphics[width=8.0cm,height=12cm]{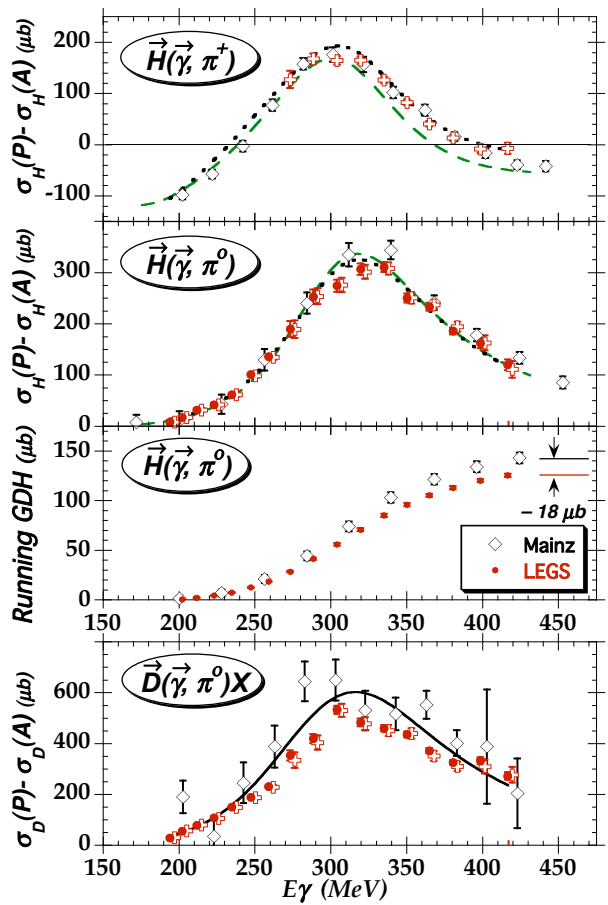}
\caption{\label{fig:tots} Total $\pi^+$ and $\pi^{\circ}$ spin-difference cross sections for 
polarized H (top two panels) and for $\pi^{\circ}$ 
production from polarized D (bottom). Open (red) crosses
result from an angle integration of the differential spin difference
($\pi^{\circ}$ crosses are shifted by +3 MeV for clarity). Solid (red) circles
result from counting $\pi$'s in the detector, using the measured
angular dependences in a simulation to correct for varying
efficiencies. Mainz results, using the latter method, are shown
as open diamonds \cite{ahrens04,ahrens06}. The $\pi^{\circ}$ contribution to the running
GDH(p) is plotted in the second to bottom panel against
the upper limit of integration. Curves are as in Fig.~\ref{fig:unpold} and Fig.~\ref{fig:angd}.}
\end{figure}

The angular distributions of the $\pi$ spin-difference cross sections
have been fitted to a Legendre expansion (solid blue curves in Fig.~\ref{fig:angd}).
The integration of these distributions are shown as 
the open (red) crosses in Fig.~\ref{fig:tots}. Our total spin difference
for $\pi^+$  from polarized H (Fig.~\ref{fig:tots}, top) is in excellent
agreement with Mainz results \cite{ahrens04}, although limited to energies
above 270 MeV by absorption in the neutron detectors surrounding
the HD target. The $\pi^{\circ}$ spin-difference is lower than the Mainz
results in the region of the $\Delta$ peak (Fig.~\ref{fig:tots}, second panel from top),
reflecting the differences in the angular distributions 

Another method of obtaining total cross sections is to simply
count pions in the full detector. This technique 
was used in Mainz experiments. All quasi-$4\pi$ detectors have
efficiencies that vary with angle, which must be corrected using
simulations. However, it is important to use accurate angular
distributions to distribute events in such simulations to avoid
biasing results, particularly when cross sections vary rapidly
with angle.  Counting neutral pions
in the SASY detector, with efficiencies corrected by simulation
using measured angular distributions, results in the solid (red) circles of
Fig.~\ref{fig:tots}. This agrees with direct integration of the
angular distributions, and has smaller uncertainties since it
avoids propagating errors from multiple background subtractions.

Systematic uncertainties on the cross section scale associated with 
target length, flux normalizations and possible geometrical differences
between the detector and the simulations are estimated at 3.5\%. Photon
beam polarizations are known to 1\%.  Systematic uncertainties on target
polarization vary between data groups. Their effect produces a 5.1\%
uncertainty in the integrated spin-difference. 
The total systematic uncertainty in GDH(p) is then 6.3\%.

The $\pi^{\circ}$ contribution to the running GDH integral for the proton is
plotted against the upper limit of integration in Fig.~\ref{fig:tots} (third panel from top).
From 200 MeV to 420 MeV, our integrated result
is 125.4 $\pm$ 1.7 $\mu$b. Integration of the Mainz data over the same
interval gives 142.9 $\pm$ 5.4 $\mu$b \cite{ahrens04}.
This difference of -17.5 $\pm$ 5.7 $\mu$b
appears to originate from a limited energy range. 
Applying this correction to the full Mainz+Bonn result, together with the
-28 $\mu$b contribution from energies below 0.2 GeV, would bring their GDH(p)
total down to 208 $\pm$ 6 (stat) $\pm$ 14 (sys) $\mu$b, where we have combined here the
systematic uncertainties from both experiments. This is to be
compared with 204 $\mu$b for the right side of eqn.~\ref{GDHeq} and removes
the need for additional canceling contributions from higher
energies to achieve agreement with the GDH(p).

The integrated spin difference for $\pi^{\circ}$ production from the deuteron
is shown in the bottom panel of Fig.~\ref{fig:tots}. These are somewhat lower than
the Mainz results of \cite{ahrens06} and considerably more precise.
The calculation of
\cite{fix} is shown as the solid curve.
While certainly in proximity to the data, further
theoretical work will be needed to address the discrepancies which are
largest in the $\pi^{\circ}$ channel (Fig.~\ref{fig:unpold} as well).

In summary, while our charged-$\pi$ data from polarized H agree with Mainz,
our $\pi^{\circ}$ results near the peak of the $\Delta$ reveal
a different angular distribution than what was assumed in
Mainz analyses. As a result, our $\pi^{\circ}$ contribution to eqn.~\ref{GDHeq} is
18 $\mu$b less than the Mainz result for H and suggests that a
high-energy Regge tail is not needed. Our results for polarized
D are lower than the trend of the Mainz data and have
considerably smaller uncertainties. The data are also
lower than recent deuteron calculations and point to
the need for additional theoretical work to understand
the GDH(D) convergence.

This work was supported by the U.S. Dept. of Energy under
Contract No. DE-AC02-98-CH10886, by the Istituto Nazionale di
Fisica Nucleare, Italy, and by the U.S. National Science Foundation.
We are indebted to Mr. F. Lincoln for his expert technical assistance.
We thank Drs. C. Commeaux,
J.-P. Didelez and G. Rouill\'{e} for their collaboration
during the early stages of the
HD target development. One of us (AMS) would like to thank
Drs. A. Fix and H. Arenh\"{o}vel for supplying their deuteron
calculations and for many clarifying discussions, as well as
Dr. T.-S. H. Lee for stimulating interactions.

\end{document}